\PassOptionsToPackage{unicode}{hyperref}
\PassOptionsToPackage{hyphens}{url}
\PassOptionsToPackage{dvipsnames,svgnames,x11names}{xcolor}
\documentclass[
  11pt,
]{article}
\usepackage{amsmath,amssymb}
\usepackage{lmodern}
\usepackage{iftex}
\ifPDFTeX
  \usepackage[T1]{fontenc}
  \usepackage[utf8]{inputenc}
  \usepackage{textcomp} 
\else 
  \usepackage{unicode-math}
  \defaultfontfeatures{Scale=MatchLowercase}
  \defaultfontfeatures[\rmfamily]{Ligatures=TeX,Scale=1}
\fi
\IfFileExists{upquote.sty}{\usepackage{upquote}}{}
\IfFileExists{microtype.sty}{
  \usepackage[]{microtype}
  \UseMicrotypeSet[protrusion]{basicmath} 
}{}
\makeatletter
\@ifundefined{KOMAClassName}{
  \IfFileExists{parskip.sty}{%
    \usepackage{parskip}
  }{
    \setlength{\parindent}{0pt}
    \setlength{\parskip}{6pt plus 2pt minus 1pt}}
}{
  \KOMAoptions{parskip=half}}
\makeatother
\usepackage{xcolor}
\usepackage{color}
\usepackage{fancyvrb}

\DefineVerbatimEnvironment{Highlighting}{Verbatim}{commandchars=\\\{\}}
\newenvironment{Shaded}{}{}

\newcommand{\CommentTok}[1]{\textcolor[rgb]{0.38,0.63,0.69}{\textit{#1}}}

\newcommand{\DecValTok}[1]{\textcolor[rgb]{0.25,0.63,0.44}{#1}}

\newcommand{\FloatTok}[1]{\textcolor[rgb]{0.25,0.63,0.44}{#1}}

\newcommand{\ImportTok}[1]{\textcolor[rgb]{0.00,0.50,0.00}{\textbf{#1}}}

\newcommand{\NormalTok}[1]{#1}
\newcommand{\OperatorTok}[1]{\textcolor[rgb]{0.40,0.40,0.40}{#1}}

\newcommand{\StringTok}[1]{\textcolor[rgb]{0.25,0.44,0.63}{#1}}
\newcommand{\VariableTok}[1]{\textcolor[rgb]{0.10,0.09,0.49}{#1}}

\usepackage{graphicx}
\makeatletter
\def\maxwidth{\ifdim\Gin@nat@width>\linewidth\linewidth\else\Gin@nat@width\fi}
\def\maxheight{\ifdim\Gin@nat@height>\textheight\textheight\else\Gin@nat@height\fi}
\makeatother
\setkeys{Gin}{width=\maxwidth,height=\maxheight,keepaspectratio}
\makeatletter
\def\fps@figure{htbp}
\makeatother
\providecommand{\tightlist}{%
  \setlength{\itemsep}{0pt}\setlength{\parskip}{0pt}}
\NewDocumentCommand\citeproctext{}{}
\NewDocumentCommand\citeproc{mm}{%
  \begingroup\def\citeproctext{#2}\cite{#1}\endgroup}
\makeatletter
 \let\@cite@ofmt\@firstofone
 \def\@biblabel#1{}
 \def\@cite#1#2{{#1\if@tempswa , #2\fi}}
\makeatother
\newlength{\cslhangindent}
\newlength{\csllabelwidth}
\newenvironment{CSLReferences}[2] 
 {\begin{list}{}{%
  \setlength{\itemindent}{0pt}
  \setlength{\leftmargin}{0pt}
  \setlength{\parsep}{0pt}
  \ifodd #1
   \setlength{\leftmargin}{\cslhangindent}
   \setlength{\itemindent}{-1\cslhangindent}
  \fi
  \setlength{\itemsep}{#2\baselineskip}}}
 {\end{list}}
\usepackage{calc}

\ifLuaTeX
\usepackage[bidi=basic]{babel}
\else
\usepackage[bidi=default]{babel}
\fi
\babelprovide[main,import]{american}

\def\languageshorthands#1{}
\usepackage[margin=1in]{geometry}
\ifLuaTeX
  \usepackage{selnolig}  
\fi
\IfFileExists{bookmark.sty}{\usepackage{bookmark}}{\usepackage{hyperref}}
\IfFileExists{xurl.sty}{\usepackage{xurl}}{} 
\hypersetup{
  pdftitle={FastBox: A lightweight Python package for rapid cosmological
21cm intensity mapping simulations},
  pdfauthor={Philip Bull, Bruno B. Bizarria, Melis O. Irfan, Geoff
Murphy},
  pdflang={en-US},
  colorlinks=true,
  linkcolor={Maroon},
  filecolor={Maroon},
  citecolor={Blue},
  urlcolor={Blue},
  pdfcreator={LaTeX via pandoc}}

\title{FastBox: A lightweight Python package for rapid cosmological 21cm
intensity mapping simulations}

\definecolor{c53baa1}{RGB}{83,186,161}
\definecolor{c202826}{RGB}{32,40,38}

\usepackage[affil-it]{authblk}
\usepackage{orcidlink}
\author[1%
  \ensuremath\mathparagraph]{Philip Bull%
    \,\orcidlink{0000-0001-5668-3101}\,%
    }
\author[2%
  ]{Bruno B. Bizarria%
    \,\orcidlink{0000-0001-7794-6599}\,%
    }
\author[3%
  ]{Melis O. Irfan%
    \,\orcidlink{0000-0003-2021-7357}\,%
    }
\author[4%
  ]{Geoff Murphy%
    \,\orcidlink{0000-0002-8186-3064}\,%
    }

\affil[1]{Jodrell Bank Centre for Astrophysics, Department of Physics \&
Astronomy, The University of Manchester, Manchester M13 9PL, UK%
  }
\affil[2]{Instituto Nacional de Pesquisas Espacias, São José dos Campos,
Brazil%
  }
\affil[3]{Institute of Astrophysics, University of Cambridge, Madingley
Road, CB3 0HA, UK%
  }
\affil[4]{Department of Physics \& Astronomy, University of the Western
Cape, Robert Sobukwe Road, Cape Town 7535, South Africa%
  }
\affil[$\mathparagraph$]{Corresponding author: %
  phil.bull@manchester.ac.uk %
}
\date{22 October 2026}

\begin{document}
\maketitle

\section{Summary}\label{summary}

\texttt{FastBox} is a Python package for generating fast, physically
realistic simulations of cosmological signals in three-dimensional
co-moving boxes, with the primary focus of its application being 21cm
intensity mapping (IM) experiments. It provides a framework for
producing cosmology-dependent Gaussian and log-normal density fields, as
well as modelling the effects of redshift-space distortions and linear
biasing, among others. Models of instrumental systematics are
incorporated, including radiometer noise and beam convolutions. Diffuse
and point source foreground models are included, along with a number of
foreground filtering strategies such as PCA, ICA, and transfer function
correction via mock signal injection. Lastly, estimators are provided
for power spectra, the two-point correlation function, and the
equilateral bispectrum. The same underlying density field can also be
Poisson-sampled into a discrete galaxy catalogue, so that 21cm and
galaxy tracers are simulated from common initial conditions and their
cross-power spectrum can be measured. Simulation grids may be cubic or
anisotropic, with independent side lengths and cell counts along each
Cartesian axis. \texttt{FastBox} is designed as a lightweight but
realistic test-bench for the development and validation of end-to-end
cosmological analysis pipelines.

\section{Statement of Need}\label{statement-of-need}

Intensity mapping is a way of making 3D maps of the redshifted 21cm
emission line from neutral hydrogen (HI). It works by measuring the
combined emission from many sources containing neutral hydrogen gas at
low angular resolution, but over a large cosmological volume. 21cm IM
surveys are a promising method for tests of cosmological models on the
largest distance scales in the post-reionisation era, utilising
instruments such as CHIME (\citeproc{ref-CHIME2025}{CHIME Collaboration
et al., 2025}), HIRAX (\citeproc{ref-Newburgh2016}{Newburgh et al.,
2016}), GBT (\citeproc{ref-2010arXiv1007.3709C}{Chang et al., 2010}),
MeerKAT (\citeproc{ref-2026arXiv261012158M}{MeerKLASS Collaboration et
al., 2026}), Parkes (\citeproc{ref-2018MNRAS.476.3382A}{Anderson et al.,
2018}), and eventually the Square Kilometre Array
(\citeproc{ref-2026aask.bookE.106W}{Wolz et al., 2026}). Foreground
contamination is a common challenge for all such IM experiments however,
with Galactic and extragalactic foreground emission being several orders
of magnitude brighter than the 21cm emission. The foregrounds are
further corrupted by frequency-dependent instrumental effects, such as
the telescope beam. This necessitates the use of statistical or
data-driven foreground removal techniques, which themselves need to be
tested on simulations to mitigate issues such as signal loss
(overfitting).

While the performance of 21cm surveys has often been predicted using
techniques such as Fisher matrix forecasts
(\citeproc{ref-2015ApJ...803...21B}{Bull et al., 2015}), these are
simplistic. Approaches such as simulated mock analyses provide more
realism, and can more accurately take into account the systematic
effects that dominate the error budget for this type of experiment.
These analyses are necessarily quite complex however, and require a way
to simulate data with the correct cosmological correlations and
foreground properties, add instrument-dependent systematic effects, and
then apply a variety of data analysis methods such as foreground
removal, signal loss correction, and power spectrum estimation in order
to understand the likely real-world performance of any given experiment.

\section{State of the Field}\label{state-of-the-field}

A number of 21cm simulation packages exist (e.g.
\citeproc{ref-2020MNRAS.499..304C}{Carucci et al., 2020};
\citeproc{ref-2026arXiv260701864C}{Chen et al., 2026};
\citeproc{ref-2026JCAP...01..019M}{Mokeddem et al., 2026};
\citeproc{ref-2022JCAP...05..002R}{Ramírez-Pérez et al., 2022}), each
making a variety of different approximations and assumptions about the
21cm field and instrumental effects. Depending on the application, these
may or may not be suitable, e.g.~for the rapid exploration of
cosmological models, testing of gridding, map-making, beam convolution,
and foreground removal effects, and different types of noise and
residual systematics. Several or even all of these features may be
needed when testing new analysis and calibration pipelines however. In
addition, these codes often provide only the simulation tools, with
basic data analysis steps left to external packages. This results in
additional complexity for many applications, as multiple codes must be
installed and interfaced to perform the semi-realistic analyses that are
needed to validate different survey designs or analysis approaches.

\section{Software Design}\label{software-design}

\texttt{FastBox} has been designed to provide a `one-stop' solution to
the problem of making semi-realistic 21cm intensity mapping data and
subsequently performing the main analysis steps needed by real-world
surveys. The package provides multiple fast cosmological simulation
options; a range of realistic foreground models; various options for
adding instrumental noise and realistic beam convolution effects; and
features such as correlated galaxy field generation, void
identification, and power spectrum and bispectrum estimation. It has few
dependencies, integrates with the widely-used \texttt{pyccl} cosmology
code (\citeproc{ref-Chisari2019}{Chisari et al., 2019}) for accurate
cosmology-dependent nonlinear power spectra, and provides a fully
self-contained suite for building and running mock data generation and
analysis pipelines. While some of the chosen methods (e.g.~log-normal
density field simulation) are approximate, they are fast and are
implemented directly within \texttt{FastBox}, rather than adding to the
list of dependencies.

\texttt{FastBox} is built around the \texttt{CosmoBox} class
(\texttt{fastbox.box}), which contains the cosmological parameters,
co-moving volume, and grid resolution of a simulation. This handles
operations such as density field generation, and brightness temperature
scaling. The box need not be cubic: both the side lengths and the number
of grid cells may be specified independently per Cartesian axis, so that
a simulated volume can be matched to the anisotropic sky-frequency
geometry of a real IM survey. \texttt{pyccl}
(\citeproc{ref-Chisari2019}{Chisari et al., 2019}) handles theoretical
cosmological computations.

The package is organised into the following submodules:

\begin{itemize}
\tightlist
\item
  \texttt{fastbox.box} -- core simulation box; density fields,
  redshift-space transforms, cubic and anisotropic grids
\item
  \texttt{fastbox.power} -- auto, cross and weighted power spectrum
  estimation, forward-modelled observational power spectra, the 1D
  two-point correlation function, and the equilateral bispectrum
\item
  \texttt{fastbox.tracers} -- HI tracer biasing, mean brightness
  temperature, mock signal generation, and Poisson-sampled galaxy
  catalogues
\item
  \texttt{fastbox.foregrounds} -- Galactic synchrotron and extragalactic
  point source foreground models
\item
  \texttt{fastbox.noise} -- radiometer noise model for multi-dish arrays
\item
  \texttt{fastbox.filters} -- foreground separation (PCA, ICA, NMF) and
  transfer function estimation
\item
  \texttt{fastbox.forecast} -- Fisher matrix forecasts for cosmological
  parameters
\item
  \texttt{fastbox.voids} -- void detection and catalogue generation
\item
  \texttt{fastbox.halos} -- halo catalogue generation
\item
  \texttt{fastbox.beams} -- beam models (Gaussian, KATBeam, Zernike)
  with FFT and direct convolutions
\item
  \texttt{fastbox.inpaint} -- Gaussian process inpainting of flagged or
  missing data
\item
  \texttt{fastbox.meerklass} -- MeerKLASS survey geometry,
  released-window loading, hit-map noise, and masked weighted PCA
\item
  \texttt{fastbox.analysis}, \texttt{fastbox.plot},
  \texttt{fastbox.utils} -- analysis helpers, plotting, and shared
  utilities
\end{itemize}

All Fourier operations use \texttt{numpy.fft}. Power spectrum
multipoles, the two-point correlation function, and the equilateral
bispectrum are estimated natively in \texttt{fastbox.power}, using
established estimators such as the Wiener-Khinchin theorem (for the
correlation function) and the Scoccimarro estimator (for the
bispectrum).

\subsection{Cross-correlation with galaxy
surveys}\label{cross-correlation-with-galaxy-surveys}

Cross-correlating an intensity map with a galaxy survey is one of the
main routes to a 21cm detection, since foreground residuals do not
correlate with galaxy positions and so add scatter to the cross spectrum
rather than bias. Simulating this requires both tracers to be drawn from
the same realisation of the density field, which
\texttt{fastbox.tracers} provides.

\texttt{GalaxyTracer} turns a density field into a discrete galaxy
catalogue by Poisson sampling, given a target number density and a field
that already carries the galaxy bias. Because the galaxies are discrete
objects rather than a continuous field, redshift-space distortions can
be applied to their positions directly, and the catalogue is then placed
back onto the simulation grid using a choice of mass-assignment scheme.

\texttt{fastbox.power} measures auto and cross spectra from the
resulting maps, optionally weighting cells to represent a survey's
selection, and corrects for the smoothing that gridding a catalogue
introduces (\citeproc{ref-Jing2005}{Jing, 2005}). Its theoretical
predictions are built on the same grid as the measurement rather than in
the continuum, so that the beam, the channel width and the gridding are
applied to the model instead of being removed from the data. Separate
models are provided for an intensity map
(\texttt{predict\_hi\_power\_spectrum}), a galaxy map
(\texttt{predict\_galaxy\_power\_spectrum}), and their cross spectrum
(\texttt{predict\_cross\_spectrum}).

\section{Usage Examples}\label{usage-examples}

The following illustrates a complete end-to-end simulation and analysis
pipeline, beginning with signal generation and ending with power
spectrum estimation and correlation function measurement.

\subsection{Generating a simulation
box}\label{generating-a-simulation-box}

A 128\(^3\)-cell box spanning \((2\,\mathrm{Gpc})^3\) at redshift
\(z = 0.8\) with an HI tracer, including a log-normal density transform
and linear plus nonlinear redshift-space distortions:

\begin{Shaded}
\begin{Highlighting}[]
\ImportTok{import}\NormalTok{ fastbox}
\ImportTok{from}\NormalTok{ fastbox.box }\ImportTok{import}\NormalTok{ CosmoBox, default\_cosmo}

\NormalTok{box }\OperatorTok{=}\NormalTok{ CosmoBox(cosmo}\OperatorTok{=}\NormalTok{default\_cosmo, box\_scale}\OperatorTok{=}\NormalTok{(}\FloatTok{2e3}\NormalTok{, }\FloatTok{2e3}\NormalTok{, }\FloatTok{2e3}\NormalTok{),}
\NormalTok{               nsamp}\OperatorTok{=}\DecValTok{128}\NormalTok{, redshift}\OperatorTok{=}\FloatTok{0.8}\NormalTok{, realise\_now}\OperatorTok{=}\VariableTok{False}\NormalTok{)}
\NormalTok{box.realise\_density()}

\CommentTok{\# Apply the HI bias and linear redshift{-}space distortions in Fourier space}
\CommentTok{\# (sigma\_v = 120 km/s accounts for nonlinear finger{-}of{-}god smearing), then}
\CommentTok{\# a log{-}normal transform so that the brightness temperature stays positive}
\NormalTok{tracer }\OperatorTok{=}\NormalTok{ fastbox.tracers.HITracer(box)}
\NormalTok{delta\_hi }\OperatorTok{=}\NormalTok{ box.linear\_rsd\_density(box.delta\_x, bias}\OperatorTok{=}\NormalTok{tracer.bias\_HI(),}
\NormalTok{                                  sigma\_v}\OperatorTok{=}\FloatTok{120.}\NormalTok{)}
\NormalTok{delta\_ln }\OperatorTok{=}\NormalTok{ box.lognormal(delta\_hi)}

\CommentTok{\# Scale by mean brightness temperature to obtain signal cube in mK}
\NormalTok{signal\_cube }\OperatorTok{=}\NormalTok{ tracer.signal\_amplitude() }\OperatorTok{*}\NormalTok{ (}\FloatTok{1.} \OperatorTok{+}\NormalTok{ delta\_ln)}
\end{Highlighting}
\end{Shaded}

\begin{figure}
\centering
\includegraphics[width=1\linewidth,height=\textheight,keepaspectratio]{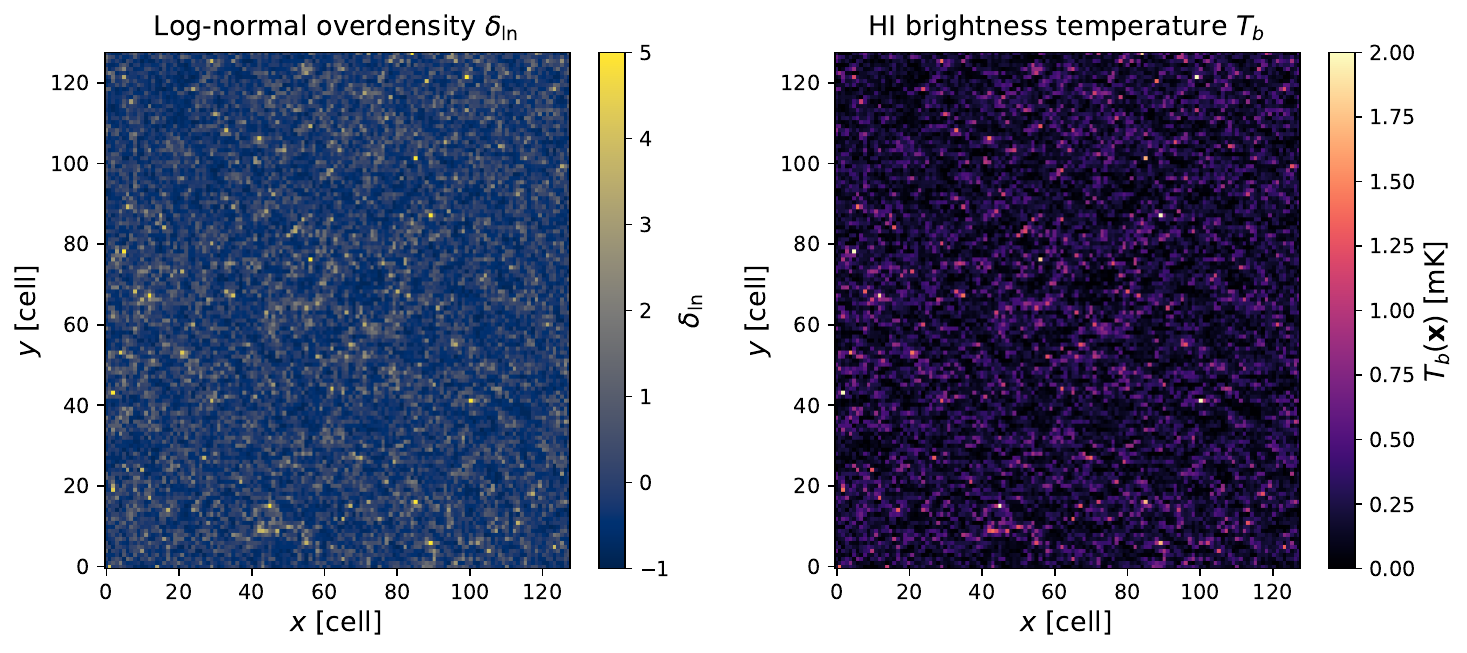}
\caption{A single frequency slice (\(z = 0.8\)) of the log-normal
overdensity field \(\delta_{\rm ln}\) (left) and the corresponding HI
brightness temperature \(T_b\) (right), generated on a \(128^3\) grid
spanning \((2\,\mathrm{Gpc})^3\).}\label{fig:field_slice}
\end{figure}

\subsection{Adding foregrounds and instrument
noise}\label{adding-foregrounds-and-instrument-noise}

Galactic synchrotron emission and extragalactic point sources are
modelled as spatially-correlated maps with power-law spectral energy
distributions, following the parameterisation of Santos et al.
(\citeproc{ref-Santos2005}{2005}):

\begin{Shaded}
\begin{Highlighting}[]
\ImportTok{from}\NormalTok{ fastbox.foregrounds }\ImportTok{import}\NormalTok{ ForegroundModel}

\NormalTok{fg }\OperatorTok{=}\NormalTok{ ForegroundModel(box)}

\CommentTok{\# Galactic synchrotron (\textasciitilde{}133 K monopole at 130 MHz)}
\NormalTok{fg\_synch\_map  }\OperatorTok{=}\NormalTok{ fg.realise\_foreground\_amp(amp}\OperatorTok{=}\FloatTok{700.}\NormalTok{, beta}\OperatorTok{={-}}\FloatTok{2.4}\NormalTok{,}
\NormalTok{                                          monopole}\OperatorTok{=}\FloatTok{133e3}\NormalTok{)}
\NormalTok{alpha\_synch   }\OperatorTok{=}\NormalTok{ fg.realise\_spectral\_index(mean\_spec\_idx}\OperatorTok{={-}}\FloatTok{2.8}\NormalTok{,}
\NormalTok{                                          std\_spec\_idx}\OperatorTok{=}\FloatTok{0.00002}\NormalTok{,}
\NormalTok{                                          smoothing\_scale}\OperatorTok{=}\FloatTok{0.1}\NormalTok{)}
\NormalTok{fg\_synch\_cube }\OperatorTok{=}\NormalTok{ fg.construct\_cube(fg\_synch\_map, alpha\_synch, freq\_ref}\OperatorTok{=}\FloatTok{130.}\NormalTok{)}

\CommentTok{\# Extragalactic point sources (\textasciitilde{}26.7 K monopole at 130 MHz)}
\NormalTok{fg\_ps\_map  }\OperatorTok{=}\NormalTok{ fg.realise\_foreground\_amp(amp}\OperatorTok{=}\FloatTok{57.}\NormalTok{, beta}\OperatorTok{={-}}\FloatTok{1.1}\NormalTok{,}
\NormalTok{                                       monopole}\OperatorTok{=}\FloatTok{26.7e3}\NormalTok{, smoothing\_scale}\OperatorTok{=}\FloatTok{0.1}\NormalTok{)}
\NormalTok{alpha\_ps   }\OperatorTok{=}\NormalTok{ fg.realise\_spectral\_index(mean\_spec\_idx}\OperatorTok{={-}}\FloatTok{2.07}\NormalTok{,}
\NormalTok{                                       std\_spec\_idx}\OperatorTok{=}\FloatTok{0.00002}\NormalTok{,}
\NormalTok{                                       smoothing\_scale}\OperatorTok{=}\FloatTok{0.1}\NormalTok{)}
\NormalTok{fg\_ps\_cube }\OperatorTok{=}\NormalTok{ fg.construct\_cube(fg\_ps\_map, alpha\_ps, freq\_ref}\OperatorTok{=}\FloatTok{130.}\NormalTok{)}

\CommentTok{\# Radiometer noise for a MeerKAT{-}like 64{-}dish deep integration}
\NormalTok{noise\_model }\OperatorTok{=}\NormalTok{ fastbox.noise.NoiseModel(box)}
\NormalTok{noise\_cube  }\OperatorTok{=}\NormalTok{ noise\_model.realise\_radiometer\_noise(Tinst}\OperatorTok{=}\FloatTok{18.}\NormalTok{, tp}\OperatorTok{=}\FloatTok{0.25}\NormalTok{,}
\NormalTok{                                                   fov}\OperatorTok{=}\FloatTok{1.}\NormalTok{, Ndish}\OperatorTok{=}\DecValTok{64}\NormalTok{)}

\NormalTok{data\_cube }\OperatorTok{=}\NormalTok{ signal\_cube }\OperatorTok{+}\NormalTok{ fg\_synch\_cube }\OperatorTok{+}\NormalTok{ fg\_ps\_cube }\OperatorTok{+}\NormalTok{ noise\_cube}
\end{Highlighting}
\end{Shaded}

\subsection{Foreground removal and transfer function
estimation}\label{foreground-removal-and-transfer-function-estimation}

PCA, ICA, and NMF foreground filters are available through a unified
interface. A bias-correction transfer function can be estimated via mock
signal injection, implementing the method introduced by Cunnington et
al. (\citeproc{ref-Cunnington2023}{2023}):

\begin{Shaded}
\begin{Highlighting}[]
\ImportTok{import}\NormalTok{ functools}

\CommentTok{\# Remove N\_fg=3 foreground modes with PCA}
\NormalTok{cleaned\_pca, U\_fg, amp\_fg }\OperatorTok{=}\NormalTok{ fastbox.filters.pca\_filter(data\_cube, nmodes}\OperatorTok{=}\DecValTok{3}\NormalTok{,}
\NormalTok{                                                        return\_filter}\OperatorTok{=}\VariableTok{True}\NormalTok{)}
\CommentTok{\# Alternative filters}
\NormalTok{cleaned\_ica, \_ }\OperatorTok{=}\NormalTok{ fastbox.filters.ica\_filter(data\_cube, nmodes}\OperatorTok{=}\DecValTok{3}\NormalTok{,}
\NormalTok{                                             return\_filter}\OperatorTok{=}\VariableTok{True}\NormalTok{)}
\NormalTok{cleaned\_nmf, \_ }\OperatorTok{=}\NormalTok{ fastbox.filters.nmf\_filter(data\_cube, nmodes}\OperatorTok{=}\DecValTok{3}\NormalTok{,}
\NormalTok{                                             return\_filter}\OperatorTok{=}\VariableTok{True}\NormalTok{)}

\CommentTok{\# Estimate PCA transfer function using 100 mock signal injections}
\NormalTok{mock\_fn }\OperatorTok{=}\NormalTok{ functools.partial(fastbox.tracers.generate\_hi\_mock,}
\NormalTok{                            cosmo}\OperatorTok{=}\NormalTok{default\_cosmo,}
\NormalTok{                            box\_scale}\OperatorTok{=}\NormalTok{(}\FloatTok{2e3}\NormalTok{, }\FloatTok{2e3}\NormalTok{, }\FloatTok{2e3}\NormalTok{),}
\NormalTok{                            nsamp}\OperatorTok{=}\DecValTok{128}\NormalTok{, redshift}\OperatorTok{=}\FloatTok{0.8}\NormalTok{)}
\NormalTok{T\_s, T\_m }\OperatorTok{=}\NormalTok{ fastbox.filters.pca\_transfer\_function(data\_cube, cleaned\_pca,}
\NormalTok{                                                  mock\_fn, box,}
\NormalTok{                                                  nmodes}\OperatorTok{=}\DecValTok{3}\NormalTok{, nmocks}\OperatorTok{=}\DecValTok{100}\NormalTok{,}
\NormalTok{                                                  nbins}\OperatorTok{=}\DecValTok{50}\NormalTok{)}
\end{Highlighting}
\end{Shaded}

\subsection{Power spectrum and correlation function
estimation}\label{power-spectrum-and-correlation-function-estimation}

\begin{Shaded}
\begin{Highlighting}[]
\ImportTok{import}\NormalTok{ numpy }\ImportTok{as}\NormalTok{ np}
\ImportTok{from}\NormalTok{ fastbox.power }\ImportTok{import}\NormalTok{ Power}

\CommentTok{\# Binned spherically{-}averaged power spectrum (transfer{-}function corrected)}
\NormalTok{k, pk, stddev }\OperatorTok{=}\NormalTok{ box.binned\_power\_spectrum(delta\_x}\OperatorTok{=}\NormalTok{cleaned\_pca, nbins}\OperatorTok{=}\DecValTok{50}\NormalTok{)}
\NormalTok{pk\_corrected  }\OperatorTok{=}\NormalTok{ pk }\OperatorTok{/}\NormalTok{ T\_m}

\CommentTok{\# Forward{-}model the prediction onto the same grid, including RSD}
\NormalTok{power }\OperatorTok{=}\NormalTok{ Power(box)}
\NormalTok{th\_k }\OperatorTok{=}\NormalTok{ np.logspace(}\OperatorTok{{-}}\DecValTok{4}\NormalTok{, }\DecValTok{1}\NormalTok{, }\DecValTok{2000}\NormalTok{)}
\NormalTok{\_, th\_pk }\OperatorTok{=}\NormalTok{ power.predict\_matter\_power\_spectrum(th\_k)}
\NormalTok{k\_im, pk\_im, \_ }\OperatorTok{=}\NormalTok{ power.predict\_hi\_power\_spectrum(th\_k, th\_pk,}
\NormalTok{                                                 bias}\OperatorTok{=}\NormalTok{tracer.bias\_HI(),}
\NormalTok{                                                 Tb}\OperatorTok{=}\NormalTok{tracer.signal\_amplitude(),}
\NormalTok{                                                 sigdeg}\OperatorTok{=}\FloatTok{0.}\NormalTok{, sigma\_nl}\OperatorTok{=}\FloatTok{120.}\NormalTok{)}

\CommentTok{\# Higher{-}order and configuration{-}space statistics}
\CommentTok{\# 1D two{-}point correlation function (Wiener{-}Khinchin, no external dependency)}
\NormalTok{r, xi }\OperatorTok{=}\NormalTok{ power.estimate\_correlation\_function(signal\_cube)}

\CommentTok{\# Equilateral bispectrum B(k, k, k) via the Scoccimarro estimator}
\NormalTok{k\_b, b\_eq, ntri }\OperatorTok{=}\NormalTok{ power.estimate\_equilateral\_bispectrum(signal\_cube, n\_bins}\OperatorTok{=}\DecValTok{8}\NormalTok{)}
\end{Highlighting}
\end{Shaded}

\begin{figure}
\centering
\includegraphics[width=1\linewidth,height=\textheight,keepaspectratio]{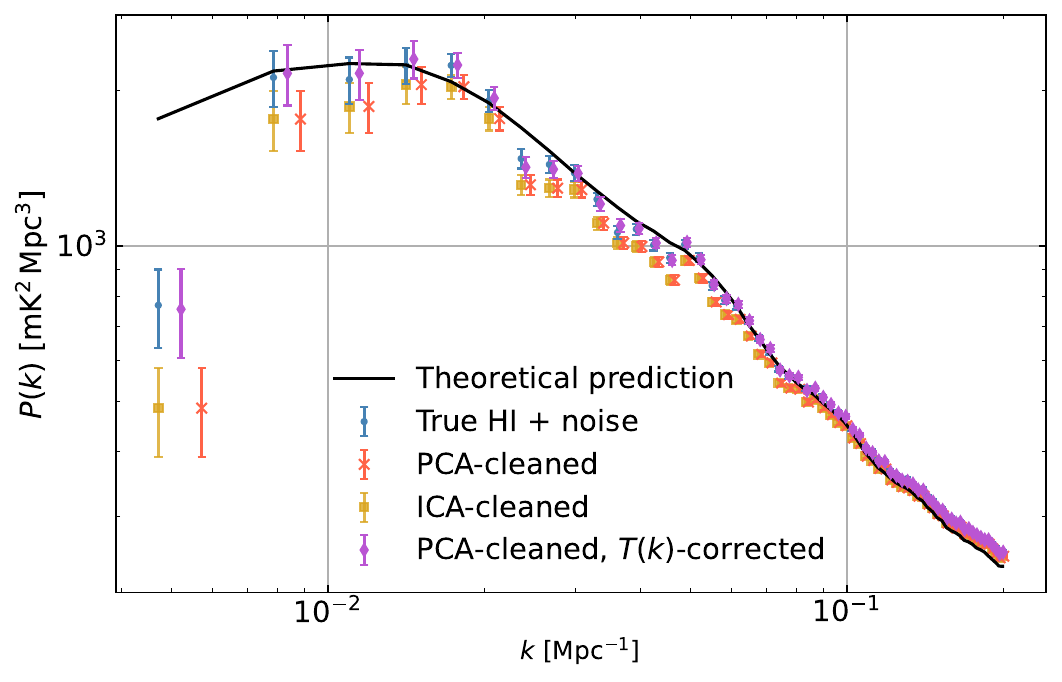}
\caption{Spherically-averaged power spectra from an end-to-end
simulation at \(z = 0.8\). The theoretical prediction (black) is
compared against the true HI signal (blue), and the signal recovered
after PCA (red) and ICA (yellow) foreground removal with
\(N_{\rm fg} = 3\) modes subtracted. The prediction is the linear
forward model \texttt{predict\_hi\_power\_spectrum}, so above
\(k \approx 0.05\,\mathrm{Mpc}^{-1}\) it falls below the true signal, to
which the log-normal transform adds small-scale power. Large-scale power
loss from foreground cleaning is visible at low \(k\). Dividing the PCA
result by the transfer function \(T(k)\), estimated here from 20 mock
signal injections, restores that lost power (purple). At high \(k\) the
corrected points lie above the true signal: \(T(k)\) compensates for
signal loss but not for the residual noise, which dominates the cleaned
map on those scales.}\label{fig:power_spectrum}
\end{figure}

\subsection{Galaxy catalogue and HI-galaxy cross
spectrum}\label{galaxy-catalogue-and-hi-galaxy-cross-spectrum}

\begin{Shaded}
\begin{Highlighting}[]
\ImportTok{from}\NormalTok{ fastbox.tracers }\ImportTok{import}\NormalTok{ GalaxyTracer}

\CommentTok{\# Galaxy overdensity from the same density field, with its own bias and the}
\CommentTok{\# same redshift{-}space distortions as the HI (GalaxyTracer expects the bias to}
\CommentTok{\# be in the field already)}
\NormalTok{b\_gal }\OperatorTok{=} \FloatTok{1.4}
\NormalTok{delta\_g }\OperatorTok{=}\NormalTok{ box.lognormal(box.linear\_rsd\_density(box.delta\_x, bias}\OperatorTok{=}\NormalTok{b\_gal,}
\NormalTok{                                               sigma\_v}\OperatorTok{=}\FloatTok{120.}\NormalTok{))}

\CommentTok{\# Poisson{-}sample a galaxy catalogue and assign it to the grid as an}
\CommentTok{\# overdensity, using cloud{-}in{-}cell weighting}
\NormalTok{gal }\OperatorTok{=}\NormalTok{ GalaxyTracer(box, vol\_density}\OperatorTok{=}\FloatTok{1e{-}3}\NormalTok{)}
\NormalTok{delta\_gal }\OperatorTok{=}\NormalTok{ gal.generate\_mesh(delta\_g, method}\OperatorTok{=}\StringTok{\textquotesingle{}CIC\textquotesingle{}}\NormalTok{, overdensity}\OperatorTok{=}\VariableTok{True}\NormalTok{)}

\CommentTok{\# HI x galaxy cross spectrum (estimate\_weighted\_power\_spectrum takes}
\CommentTok{\# per{-}cell weights instead)}
\NormalTok{k, pk\_cross, sigma\_cross }\OperatorTok{=}\NormalTok{ power.estimate\_power\_spectrum(cleaned\_pca,}
\NormalTok{                                                         delta\_gal)}

\CommentTok{\# Forward{-}model the prediction onto the same grid, with the CIC window applied}
\CommentTok{\# to the model rather than deconvolved from the data. These maps have no beam;}
\CommentTok{\# for beam{-}convolved maps, set sigdeg to the beam width in degrees}
\NormalTok{k\_cc, pk\_cc, \_ }\OperatorTok{=}\NormalTok{ power.predict\_cross\_spectrum(th\_k, th\_pk,}
\NormalTok{                                              bias\_HI}\OperatorTok{=}\NormalTok{tracer.bias\_HI(),}
\NormalTok{                                              bias\_gal}\OperatorTok{=}\NormalTok{b\_gal,}
\NormalTok{                                              Tb}\OperatorTok{=}\NormalTok{tracer.signal\_amplitude(),}
\NormalTok{                                              sigdeg}\OperatorTok{=}\FloatTok{0.}\NormalTok{, MAS}\OperatorTok{=}\StringTok{\textquotesingle{}CIC\textquotesingle{}}\NormalTok{,}
\NormalTok{                                              sigma\_nl}\OperatorTok{=}\FloatTok{120.}\NormalTok{)}
\end{Highlighting}
\end{Shaded}

The forward models are linear in the density field. On large scales the
measured cross spectrum therefore falls below the model due to PCA
signal loss, which the transfer function above corrects; on small scales
it rises above the model, because the log-normal transform adds power
that a linear model does not include.

\section{Research Impact}\label{research-impact}

\texttt{FastBox} simulated 21cm data cubes were used in Irfan \& Bull
(\citeproc{ref-Irfanux26Bull2021}{2021}), where kernel principal
component analysis was tested as an alternative to standard principal
component analysis (PCA) foreground removal techniques. KPCA was found
to improve recovery at intermediate-to-large cosmological scales. Irfan
(\citeproc{ref-Irfan2023}{2023}) generated mock 21cm cubes to test PCA
foreground removal in four cases of Galactic synchrotron spectral index
models, including a high-resolution model they generated via a
convolutional neural network (CNN). Irfan et al.
(\citeproc{ref-Irfan2024}{2024}) used \texttt{FastBox} mock signal and
foreground cubes as inputs for pipelines focused on the removal of
correlated noise.

Engelbrecht et al. (\citeproc{ref-Engelbrecht2025}{2025}) used mock
\texttt{FastBox} signal cubes to validate their Radio Navigation
Satellite System Radio Frequency Interference (RFI) model against
MeerKAT data. Murphy et al. (\citeproc{ref-Murphy2026}{2026}) used mock
signal and foreground datacubes to validate the statistical separation
of 21cm signal and foregrounds in an IM-like experiment using Gibbs
sampling and Gaussian Constrained Realisations.

\section{AI Usage Disclosure}\label{ai-usage-disclosure}

Generative AI was used to create an initial outline of this manuscript
in-keeping with JOSS requirements, which the authors thereafter verified
and further iterated on. The transfer function correction implements the
mock signal injection method of Cunnington et al.
(\citeproc{ref-Cunnington2023}{2023}), and is not a new method. The
implementation contributed here was written independently for Murphy et
al. (\citeproc{ref-Murphy2026}{2026}); generative AI was then used to
adapt it to \texttt{FastBox}'s conventions, for example the inclusion of
the docstring. Lastly, AI was used to perform additional tests on the
code base, with any bug and error reports being verified for validity by
the authors before implementation.

\section{Acknowledgements}\label{acknowledgements}

B.B.B. acknowledges resources received from Fundação de Amparo à
Pesquisa do Estado de São Paulo (FAPESP) under process 2022/16749-0.
This result is part of a project that has received funding from the
European Research Council (ERC) under the European Union's Horizon 2020
research and innovation programme (Grant agreement No.~948764; PB).
G.G.M. acknowledges the financial assistance of the South African Radio
Astronomy Observatory (SARAO) (www.sarao.ac.za).

\section*{References}\label{references}
\addcontentsline{toc}{section}{References}

\protect\phantomsection\label{refs}


\begin{CSLReferences}{1}{0}
\bibitem[\citeproctext]{ref-2018MNRAS.476.3382A}
Anderson, C. J., Luciw, N. J., Li, Y.-C., Kuo, C. Y., Yadav, J., Masui,
K. W., Chang, T.-C., Chen, X., Oppermann, N., Liao, Y.-W., Pen, U.-L.,
Price, D. C., Staveley-Smith, L., Switzer, E. R., Timbie, P. T., \&
Wolz, L. (2018). {Low-amplitude clustering in low-redshift 21-cm
intensity maps cross-correlated with 2dF galaxy densities}.
\emph{MNRAS}, \emph{476}(3), 3382--3392.
\url{https://doi.org/10.1093/mnras/sty346}

\bibitem[\citeproctext]{ref-2015ApJ...803...21B}
Bull, P., Ferreira, P. G., Patel, P., \& Santos, M. G. (2015).
{Late-time Cosmology with 21 cm Intensity Mapping Experiments}.
\emph{ApJ}, \emph{803}(1), 21.
\url{https://doi.org/10.1088/0004-637X/803/1/21}

\bibitem[\citeproctext]{ref-2020MNRAS.499..304C}
Carucci, I. P., Irfan, M. O., \& Bobin, J. (2020). {Recovery of 21-cm
intensity maps with sparse component separation}. \emph{MNRAS},
\emph{499}(1), 304--319. \url{https://doi.org/10.1093/mnras/staa2854}

\bibitem[\citeproctext]{ref-2010arXiv1007.3709C}
Chang, T.-C., Pen, U.-L., Bandura, K., \& Peterson, J. B. (2010).
{Hydrogen 21-cm Intensity Mapping at redshift 0.8}. \emph{arXiv
e-Prints}, arXiv:1007.3709.
\url{https://doi.org/10.48550/arXiv.1007.3709}

\bibitem[\citeproctext]{ref-2026arXiv260701864C}
Chen, Z., Cunnington, S., Tassie, D., Pourtsidou, A., Wolz, L., Autieri,
G., Barberi-Squarotti, M., Bernal, J. L., Bull, P., Camera, S., Carucci,
I. P., Engelbrecht, B., Fonseca, J., Fornazier, K., Grainge, K., Han,
J., Hu, W., Irfan, M. O., Kittiwisit, P., \ldots{} Zhao, B. (2026).
{meer21cm: an Analysis Pipeline and Comprehensive Toolkit for HI
Intensity Mapping}. \emph{arXiv e-Prints}, arXiv:2607.01864.
\url{https://doi.org/10.48550/arXiv.2607.01864}

\bibitem[\citeproctext]{ref-CHIME2025}
CHIME Collaboration, Amiri, M., Bandura, K., Chakraborty, A., Cliche,
J.-F., Dobbs, M., Foreman, S., Gray, L., Halpern, M., Hill, A. S.,
Hinshaw, G., Höfer, C., Joseph, A., Kruger, N., Landecker, T. L., van
Lieshout, R., MacEachern, J., Masui, K. W., Mena-Parra, J., \ldots{}
Wulf, D. (2025). {Detection of the Cosmological 21 cm Signal in
Auto-correlation at z \raisebox{-0.5ex}\textasciitilde1 with the
Canadian Hydrogen Intensity Mapping Experiment}. \emph{arXiv e-Prints},
arXiv:2511.19620. \url{https://doi.org/10.48550/arXiv.2511.19620}

\bibitem[\citeproctext]{ref-Chisari2019}
Chisari, N. E., Alonso, D., Krause, E., Leonard, C. D., Bull, P., Neveu,
J., Villarreal, A., Singh, S., McClintock, T., Ellison, J., Du, Z.,
Zuntz, J., Mead, A., Joudaki, S., Lorenz, C. S., Tröster, T., Sanchez,
J., Lanusse, F., Ishak, M., \ldots{} LSST Dark Energy Science
Collaboration. (2019). Core cosmology library: Precision cosmological
predictions for {LSST}. \emph{ApJS}, \emph{242}(1), 2.
\url{https://doi.org/10.3847/1538-4365/ab1658}

\bibitem[\citeproctext]{ref-Cunnington2023}
Cunnington, S., Wolz, L., Bull, P., Carucci, I. P., Grainge, K., Irfan,
M. O., Li, Y., Pourtsidou, A., Santos, M. G., Spinelli, M., \& Wang, J.
(2023). The foreground transfer function for {HI} intensity mapping
signal reconstruction: {MeerKLASS} and precision cosmology applications.
\emph{MNRAS}, \emph{523}(2), 2453--2477.
\url{https://doi.org/10.1093/mnras/stad1567}

\bibitem[\citeproctext]{ref-Engelbrecht2025}
Engelbrecht, B. N., Santos, M. G., Fonseca, J., Li, Y., Wang, J., Irfan,
M. O., Harper, S. E., Grainge, K., Bull, P., Carucci, I. P., Cunnington,
S., Pourtsidou, A., Spinelli, M., \& Wolz, L. (2025). {Radio frequency
interference from radio navigation satellite systems: simulations and
comparison to MeerKAT single-dish data}. \emph{MNRAS}, \emph{536}(1),
1035--1055. \url{https://doi.org/10.1093/mnras/stae2649}

\bibitem[\citeproctext]{ref-Irfan2023}
Irfan, M. O. (2023). {Simulating a full-sky high resolution Galactic
synchrotron spectral index map using neural networks}. \emph{MNRAS},
\emph{520}(4), 6070--6082. \url{https://doi.org/10.1093/mnras/stad526}

\bibitem[\citeproctext]{ref-Irfanux26Bull2021}
Irfan, M. O., \& Bull, P. (2021). {Cleaning foregrounds from single-dish
21 cm intensity maps with Kernel principal component analysis}.
\emph{MNRAS}, \emph{508}(3), 3551--3568.
\url{https://doi.org/10.1093/mnras/stab2855}

\bibitem[\citeproctext]{ref-Irfan2024}
Irfan, M. O., Li, Y., Santos, M. G., Bull, P., Gu, J., Cunnington, S.,
Grainge, K., \& Wang, J. (2024). {Mitigating the effect of 1/f noise on
the detection of the H I intensity mapping power spectrum from
single-dish measurements}. \emph{MNRAS}, \emph{527}(3), 4717--4729.
\url{https://doi.org/10.1093/mnras/stad3457}

\bibitem[\citeproctext]{ref-Jing2005}
Jing, Y. P. (2005). Correcting for the alias effect when measuring the
power spectrum using a fast fourier transform. \emph{The Astrophysical
Journal}, \emph{620}(2), 559--563. \url{https://doi.org/10.1086/427087}

\bibitem[\citeproctext]{ref-2026arXiv261012158M}
MeerKLASS Collaboration, Barberi-Squarotti, M., Bernal, J. L., Bull, P.,
Camera, S., Carucci, I. P., Chen, Z., Cunnington, S., Engelbrecht, B.
N., Fonseca, J., Grainge, K., Hu, W., Irfan, M. O., Li, Y., Santos, M.
G., Spinelli, M., Wang, J., Witzemann, A., \& Wolz, L. (2026). {HI
intensity mapping with MeerKAT: A measurement of the 21-cm auto-spectrum
on cosmological scales}. \emph{arXiv e-Prints}, arXiv:2610.12158.
\url{https://doi.org/10.48550/arXiv.2610.12158}

\bibitem[\citeproctext]{ref-2026JCAP...01..019M}
Mokeddem, R., Bizarria, B. B., Zhang, J., Hipólito-Ricaldi, W. S.,
Wuensche, C. A., Abdalla, E., Abdalla, F. B., Queiroz, A. R., Villela,
T., Wang, B., Feng, C., Gurjão, E. C., \& Marins, A. (2026).
{Cosmological remapping for efficient generation of 21 cm intensity
mapping mocks}. \emph{JCAP}, \emph{2026}(1), 019.
\url{https://doi.org/10.1088/1475-7516/2026/01/019}

\bibitem[\citeproctext]{ref-Murphy2026}
Murphy, G. G., Bull, P., Santos, M. G., Zhang, Z., \& Cunnington, S.
(2026). {Bayesian component separation and power spectrum estimation for
21 cm intensity mapping data cubes}. \emph{arXiv e-Prints},
arXiv:2604.26890. \url{https://doi.org/10.48550/arXiv.2604.26890}

\bibitem[\citeproctext]{ref-Newburgh2016}
Newburgh, L. B., Bandura, K., Bucher, M. A., Chang, T.-C., Chiang, H.
C., Cliche, J. F., Davé, R., Dobbs, M., Clarkson, C., Ganga, K. M.,
Gogo, T., Gumba, A., Gupta, N., Hilton, M., Johnstone, B., Karastergiou,
A., Kunz, M., Lokhorst, D., Maartens, R., \ldots{} Witzemann, A. (2016).
{HIRAX}: A probe of dark energy and radio transients. \emph{Proc. SPIE},
\emph{9906}, 99065X. \url{https://doi.org/10.1117/12.2234286}

\bibitem[\citeproctext]{ref-2022JCAP...05..002R}
Ramírez-Pérez, C., Sanchez, J., Alonso, D., \& Font-Ribera, A. (2022).
{CoLoRe: fast cosmological realisations over large volumes with multiple
tracers}. \emph{JCAP}, \emph{2022}(5), 002.
\url{https://doi.org/10.1088/1475-7516/2022/05/002}

\bibitem[\citeproctext]{ref-Santos2005}
Santos, M. G., Cooray, A., \& Knox, L. (2005). Multifrequency analysis
of 21 centimeter fluctuations from the era of reionization. \emph{ApJ},
\emph{625}(2), 575--587. \url{https://doi.org/10.1086/429857}

\bibitem[\citeproctext]{ref-2026aask.bookE.106W}
Wolz, L., Autieri, G., Bernal, J. L., Berti, M., Bull, P., Camera, S.,
Carucci, I., Chen, Z., Cunnington, S., Fonseca, J., Grainge, K., Han,
J., Hu, W., Karagiannis, D., Santos, M. G., Spinelli, M.,
Randrianjanahary, L. F., Uhlemann, C., Viel, M., \ldots{} Wang, J.
(2026). {Cosmology with HI Intensity Mapping}. In A. Bonaldi, T. L.
Bourke, P. Hartley, T. An, M. Audard, O. Bayandina, N. Bellomo, E.
Bianchi, M. Burgay, J. Callingham, S. Camera, V. Casasola, V. Cuciti, P.
Cole, N. Gupta, C. L. Hale, I. Harrison, J. Hessels, T. Huege, \ldots{}
P. Zucca (Eds.), \emph{Advancing astrophysics with the {SKA II}} (p.
106). \url{https://doi.org/10.48550/arXiv.2607.03259}

\end{CSLReferences}
\end{document}